\documentclass[12pt]{article}
\usepackage{hyperref}
\usepackage[superscript]{cite}
\makeatletter
\renewcommand\@biblabel[1]{#1.)}
\makeatother
\begin{document}

\title{On the Nature of Science}
 
\author{B.K.~Jennings\footnote{e-mail: jennings@triumf.ca}\\TRIUMF,
4004 Wesbrook Mall, Vancouver, BC, V6T 2A3}

\maketitle

\begin{abstract}
A 21st century view of the nature of science is presented. It attempts
to show how a consistent description of science and scientific
progress can be given. Science advances through a sequence of models
with progressively greater predictive power.  The philosophical and
metaphysical implications of the models change in unpredictable ways
as the predictive power increases. The view of science arrived at is
one based on instrumentalism. Philosophical realism can only be
recovered by a subtle use of Occam's razor. Error control is seen to
be essential to scientific progress.  The nature of the difference
between science and religion is explored.
\end{abstract}

\section{Introduction}

The idea that empirical knowledge, {\em i.e.} knowledge obtained
through observations, is unreliable has a long history going back to
the ancient Greeks. They were very good at philosophy and mathematics
but made rather little progress in the empirical sciences. Aristotle's
philosophy is still studied but most people do not know that there is
an Aristotelian physics let alone what it is.  The modern study of
physics or indeed of science goes back rather to Galileo and Newton
with an honorable mention to Francis Bacon for his formulation of
scientific induction\cite{Bacon}.  This development continued up to
Einstein's 1905 papers. The 1900 understanding of science is nicely
described in ``A pilgrimage to Popper'' by Adam Gopnik in The New
Yorker\cite{Gopnik}:
\begin{quote}
Scientists, it seemed clear, began with careful observations,
cautiously proceeded to a tentative hypothesis, progressed to more
secure but still provisional theories, and only in the end achieved,
after a long process of verification, the security of permanent
laws. Newton saw the apple fall, hypothesized that it had fallen at
one speed rather than another for a reason, theorized that there might
be an attraction between all bodies with mass, and then, at long last,
arrived at a law of gravitation to explain everything. This
``observation up'' or ``apple down'' picture of how science works was
so widespread that it defined what we mean by science: when Sherlock
Holmes says that he never theorizes in advance of the facts, he is
explaining why he can be called a scientific detective. Various
thinkers poked holes in this picture, but generally their point was
that, while the program was right, it was harder to do than it
appeared.
\end{quote}
The problem\cite{Hume} with this description of science was pointed
out by David Hume (1711-1776); namely that it was impossible to deduce
from a finite series of observations a generalization that applies to
all. For example, I have seen many crows. All of them are black. Does
this make the statement ``All crows are black.'' a fact? Indeed not,
there may still be white crows. I have read that albino crows exist
but have never seen one. While others noted Hume's objection it was
regarded as something to disprove. Furthermore, they noted it was the
job of philosophy to explain what was wrong with Hume's argument.
Kant's ``Critique of Pure Reason''\cite{Kant} was, in part, a response
to Hume.  Einstein's 1905 papers which showed the shortcomings of
classical mechanics also made Hume's arguments much more immediate and
highlighted problems in Kant's response.  Despite the great success
and great predictive power of classical mechanics it was shown to have
a limited region of validity.  The vast number of observations that
confirmed classical mechanics did not prove it universally
correct. Indeed, as Hume stressed, no number of observations can prove
a model correct.

A new understanding of science and how it worked was needed. A major
contribution to this new understanding\cite{Popper} was made by Sir
Karl Popper. His first point was that {\em all} empirical knowledge,
scientific models, and scientific laws are only tentative. It is
impossible to scientifically prove anything. It is impossible to
empirically establish any fact. This is actually a slight over
statement; there are three empirical facts.  Descartes in his
``Meditation 2'' has the famous statement ``I think, therefore I am''
or in Latin, ``cogito ergo sum''. Thus we have the two facts: ``I
think'' and its corollary ``I am''. I would add ``I observe'' to this
list of facts (sum, cogito, inspecto). Beyond that everything depends
on assumptions that can not be rigorously proven.
  
Popper's second point was that although models could not be verified
they could be falsified. Actually the idea of using false predictions
as a filter has a long history, much predating Popper or even the
scientific revolution. Deuteronomy 18:21-22 says false prophets can be
distinguished because their predictions do not come true.  The basic
idea of falsification of models is this: While seeing any number of
black crows does not prove all crows are black, seeing one white crow
disproves it. Thus science proceeds not by proving models correct but
by discarding false ones or improving incomplete ones. While the
concept of falsification is widely used in science it is also not
straightforward. Is that white bird I see really a crow? Have I
misidentified a pigeon or is it an altogether new species?  The
problem is that no theory or model exists in isolation but is always
supported by subsidiary theories and models.  Thus any test is not
just of one model but of all the subsidiary ones simultaneously.  This
has led to claims that theories are not even falsifiable. In turn this
leads to the postmodern idea that all theories or models are equally
valid. Thus we have returned to where we have started with empirically
derived knowledge in dispute.

In the next section I will present a post-postmodern view; namely
that Popper is essentially correct but with predictive power replacing
naive falsification. The distinguishing feature of any scientific
model is that it must make predictions that can be tested against
careful reproducible observations. 
 
\section{The Scientific Method}

\subsection{Observations}

How do we get around the problem of Descartes or Hume's skepticism? To
a large extent we can not: except for the three facts mentioned above
all empirical knowledge is tentative. Following Popper, we can make a
model and see how well it describes past observations and predicts
future observations. The fundamental idea of model building
(hypothesizing) and testing against observation actually goes back
much farther, at least to 1267 and Roger Bacon's {\em Opus majus}. It
even predates the idea of scientific induction. The first model
assumption we have to make is that it is interesting or useful to
study the information our senses provide {\em i.e.}  observations.  It
is not necessary to assume the sensory information corresponds to an
objective reality only that the studying of it is interesting or
useful.

That observations should be studied, although widely ingrained in
Western thought, is a relatively new idea. The clash between Galileo
and the church was primarily one between received wisdom based on the
Bible and Aristotle versus observations, especially those through the
telescope. Sunspots, the phases of Venus, mountains and valleys on the
moon, and the moons of Jupiter all challenged the received wisdom of
the perfect unchanging heavens with the earth at its center. The
ultimate question was which do you believe: your eyes or the sacred 
texts as interpreted by the religious authorities.  The wisdom of the
time was that observations were inherently untrustworthy {\em i.e.}
you should not trust your eyes but rather the authorities. Ultimately
the scientific method, which is being considered here, allows useful
information to be extracted from observations.

Observations are any sensory input. At the simplest level they are
direct sensory input. At the next level we use instruments to augment
our senses, for example Galileo's telescope. However, observations are
not just the passive observation of what nature presents to us. We
also do controlled experiments that actively manipulate nature in
order to isolate different effects and test specific aspects of the
models. The experiments are designed to maximize the information that
can be obtained while eliminating uninteresting and spurious
effects. An extreme example is the ATLAS detector at the Large Hadron
Collider (CERN, Geneva, Switzerland) which is about the size of a five
story building and involves approximately 1500 physicists from 36
countries. As the instrument gets larger and more complex the
observations become more model dependent. A model of the apparatus is
needed or even models of parts of the apparatus. However, even when
the apparatus is as simple as Galileo's telescope, a model is needed
to understand its behavior. One of the attacks against Galileo was
that he was seeing an artifact of the telescope. Thus we have the idea
that observations are not freestanding but take on their meaning
within the context of a model or models.

\subsection{Models}

Having briefly considered what an observation is we now turn in more
detail to defining the concept of a model. A model is any theoretical
construct used to describe or predict observations. Thus classical
mechanics, evolution, special creation, and road maps are models in
the present sense of the word. To be science, a model must be
internally consistent and logical. The word model has been chosen
following the practice in subatomic physics with ``the standard model
of particle physics'' or ``the shell model''. It implies that the
model is not reality but, at best, an image of it.

Consider a road map. It is a very simple example of a model. It
provides predictions about where the streets are and what their names
are. Thus we encounter Vine Street, Yew Street and Arbutus Street when
we drive down 16th Avenue; the same as the map tells us (See
maps.google.ca for Vancouver, BC, Canada). Now just because the street
sign does not agree with the map does not mean the map is wrong ---
the street sign could be incorrect. This illustrates the problem with
naive falsification. There are two models in play, one the map and the
other that the street sign correctly identifies the street. While
normally the street sign is correct, city work crews have been known
to make mistakes. Or part of the sign has been destroyed so ``Vine
Street'' becomes ``ne Street''. The map and street signs could also be
in different languages. Thus one observation, the name on the street
sign, does not necessarily indicate that the map is incorrect. However
if the map consistently has the wrong names and streets in the wrong
place, it is not useful. The map must have predictive power to be
useful.

A much more sophisticated example of a model is Kuhn's
paradigm\cite{Kuhn}.  A paradigm is the set of interlocking
assumptions and methodologies that define a field of study.  It
provides the foundation for all work in the field and a common
language for discourse. While observations exist independant of the
paradigm, their interpretation depends on the paradigm. No natural
history or any set observation can be interpreted or even usefully
discussed in the absence of the intertwined theoretical and
methodological system provided by a paradigm. Paradigms help
scientific communities to bound their discipline and create avenues of
inquiry.  They determine which are the important question to be
considered (for examples see the discussion in Sec.~\ref{sec:gdt} on
the shape of the earth).  Students and new practitioners learn the
paradigm in order to become effective members of the community. The
paradigm can also act to prevent progress when members of a community
are too committed to their current set of models.  Hence outsiders,
like Wegener\cite{Wegener} who proposed continental drift, are
sometimes at an advantage in seeing past the current models to propose
striking new approaches.  However, scientific progress cannot take
place without the framework provided by a model or paradigm.

A theory, defined by the American Heritage Dictionary\cite{AHD} as
``Systematically organized knowledge applicable in a relatively wide
variety of circumstances, especially a system of assumptions accepted
principles, and rules of procedure devised to analyze, predict, or
otherwise explain the nature or behavior of a specified set of
phenomena'', is another example of the model concept. A theory is
closely related to Kuhn's paradigm but more restrictive, not covering
methodologies. The special theory of relativity or the theory of
evolution are models in the current sense of the word. The word theory
has an unfortunate alternate meaning of hunch or conjecture so even
well supported theories like special relativity or evolution tend to
be tainted by the use of the word. As an example of this see the
Answer\cite{DoverA} to the Complaint in the Kitzmiller v. Dover Area
School District court case on the role of evolution in the classroom
where it is stated that the ``Defendants deny that the term theory, as
used in science, has a distinct meaning and does not suggest
uncertainty, doubt or speculation.'' When speaking to people with such
a limited understanding, the term ``theory'' should be avoided to
reduce confusion. It is partly for this reason that the present work
uses the term ``model''.

Hooke's Law on the force required to stretch or compress a spring,
Boyle's Law on relation between the pressure and volume of an ideal
gas, or the OZI rule\cite{Okubo} on the decay properties of
hyperons are also examples of models but with a more limited scope
than either theories or paradigms.

Historically there was a change in the nature of scientific models
with Newton. Ptolemy, Copernicus and Newton all developed models that
correctly predicted planetary motion. While it is sometimes claimed
that the Ptolemaic and Copernican models were only descriptive in
contrast to Newton's which was explanatory, it is more precise to say
that Newton introduced a higher level of abstraction using ideas
farther removed from the observations. He used the concept of force,
universal gravitation and his laws of motion while Ptolemy and
Copernicus just worked with the positions of the planets. The
abstraction level in physics increased significantly farther with
Maxwell's equations for electromagnetism and with the development of
statistical mechanics. The ether was an attempt to make the rather
abstract Maxwell's equations more concrete. When first introduced
molecules, essential in understanding chemistry and statistical
mechanics, were also a very abstract concept and, like quarks, they
were considered to be just a convenient theoretical construct. They
were rejected by many scientists including Mach. This rejection was
probably a contributing factor in Boltzmann's suicide. Quantum
mechanics and quantum field theory are progressively more
abstract. The many worlds interpretation of quantum mechanics
(discussed in sec.~\ref{sec:se}) is an attempt to make the very
abstract concepts in quantum mechanics more concrete. Like Newton in
physics, Darwin increased the level of abstraction in biology with the
idea of evolution and especially natural selection. Even his idea of
heredity was rather abstract for the time. Error control and
estimation also involves rather abstract concepts. An increase in the
level of abstraction is common when science advances. This has the
twin effects of making the models more powerful but also less
comprehensible.

N.T. Wright in his defense of the historicity of Jesus against
post-modernism\cite{Wright} uses the term ``explanatory story''. This
is largely synonymous with model. ``Explanatory story'' also has a
similar connotation to Kuhn's paradigm although in that case
``controlling narrative'' might be a better terminology.  If you
replace ``explanatory story'' with ``model'', Wright's description of
extracting information from the historical record is essentially the
same as the approach to scientific model building given here (see in
particular the last paragraph on page 37).  In his case the
``observations'' are historical records including the Bible. However,
the approach in the present work leads to instrumentalism rather than
the critical realism\cite{Polanyi} Wright advocates.

While in order to be useful in science a model must be logical and
internally consistent it does not have to be ``reasonable''. When
there is major change of models, what Kuhn would call a paradigm
shift, the new model is, by definition, unreasonable or even aberrant
in the context of the old model. For example, the action at a distance
in Newton's law of gravity was considered unreasonable by Leibniz,
Huygens and even Newton himself. Similarly, several of the founders of
quantum mechanics --- Planck, Bohr, Schr\"odinger, Einstein ---
considered quantum mechanics unreasonable. Planck apologized for
Einstein taking his quantum idea seriously. Niels Bohr
said\cite{quantum} ``Anyone who is not shocked by quantum theory has
not understood a single word.''. Erwin Schr\"odinger said\cite{Schr}:
``If we are going to stick to this damned quantum-jumping, then I
regret that I ever had anything to do with quantum theory.'' Einstein
was a critic of quantum theory and his most cited paper\cite{Einstein}
was an attempt to show that quantum mechanics must be incomplete. It
introduced what is now known as the Einstein, Podolsky, Rosen (EPR)
paradox and quantum entanglement. Unfortunately for Einstein,
observations agree with quantum mechanics and it is reality that is
unreasonable. The theory of continental drift first proposed by Alfred
Wegener in 1915\cite{Wegener} was rejected as unreasonable --- indeed
it was. How could continents plow through the oceans? It was not until
the model of plate tectonics was developed in the 1960's that
continental drift was accepted. The idea in evolution that complex
structures and interdependencies can arise through natural selection
is regarded by many as unreasonable.  String theory, the currently
proposed model of everything, is unreasonable with at least eleven
dimensions instead of the perceived four. However, it is predictive
power not reasonableness that is the final arbiter in science. Even
very unreasonable models, like Newtonian gravity, quantum mechanics,
continental drift, evolution or string theory, must be accepted {\em
if but only if} they correctly predict observations.  Simply because a
model is unreasonable does not mean it is correct. Most unreasonable
models are simply that --- unreasonable.

Finally in the discussion of models I quote\cite{quantuma,quantum}
Niels Bohr again: ``There is no quantum world. There is only an
abstract physical description. It is wrong to think that the task of
physics is to find out how nature is. Physics concerns what we can say
about nature.''  Models are what we can say about nature and are the
essence of science.

\subsection{Predictive Power}

We now have some idea of what is meant by observations and models.
Next we turn to the idea of falsification stressed by Popper. Popper's
basic idea is correct, models cannot be proven but still can be tested
by comparison with observations. However, rather than strict
falsification we judge a model by its predictive power. Popper
recognized that models which made more predictions were also more
easily falsified by having some of the prediction proven incorrect. In
spite of this, such models are preferred as having more information
content. This is moving in the direction of predictive
power. Predictive power simply refers to the ability of a model to
describe past observations and especially predict new ones. The
greater the ability to make correct predictions and not make false
ones the greater its predictive power. In some cases the predictive
power is quantitative as in $\chi^2$ fits to data but frequently it is
more qualitative. However, the more precise the prediction the
better. The new predictions must be made without additional
assumptions. In general what is important is the number of predictions
for a given number of assumptions. Increasing the number of
assumptions should only be done if there is a significant increase in
the number of correct predictions that can be made. Although models
can not be proven correct, or strictly falsified, they can be judged
and compared.

Older theories are falsified, or to be more precise, they are shown to
have a limited range of validity usually by a series of
observations. The Michelson-Morley experiment\cite{Michelson} on the
speed of light in moving frames and experiments on atomic structure
were inconsistent with the predictions of classical mechanics. This
led respectively to special relativity and quantum mechanics. The
models that replaced classical mechanics had greater predictive power,
keeping the success of the previous models while also describing the
new observations.  Special Relativity has more predictive power than
classical mechanics describing both slowly and quickly moving
objects. Quantum mechanics replaced Newton's Laws because of its
greater predictive power --- it described microscopic as well as
macroscopic systems.  Following the Correspondence Principle as stated
by Niels Bohr, quantum mechanics must reduce to and indeed does reduce
to classical mechanics in those instances when classical mechanics
provides a good description of the observations.  Evolution replaced
animals reproducing after their kind because it has more predictive
power. It makes correct predictions about fossils and the distribution
of species that are outside the scope of the
animals-reproducing-after-their-kind model.

A single false prediction usually does not doom a model. Like the
example given above with the road map, where the road sign was wrong
not the map, the problem may be with the observation. When a model
disagrees with an observation something has been falsified, it is just
not clear what. It could be the model (the map), the observation (the
street sign wrong or incorrectly read) or the model misapplied (we are
on 41st Avenue not 16th). The model may also need to be slightly
modified (a street added since the map was made) to accommodate the
new finding.  However if separate modifications for a wide of range of
new observation are needed the model loses its ability to predict and
must be improved or rejected.

A prime example of one experiment (actually a series of experiments by
the same person) not being sufficient to falsify a model is the Dayton
Miller\cite{Miller} followup to the Michelson-Morley
experiment. Miller, in a series of experiments, found an effect on the
speed of light due to the earth's motion. If accepted this would have
falsified special relativity. While this experiment was widely
discussed at the time, Einstein largely dismissed it. In the end it
was the Michelson-Morley experimental result that could be
reproduced\cite{Brillet} not the Miller result.  Experimental results
must be reproducible to be useful.  As discussed in the next section
most exciting new results are wrong. It was the Michelson-Morley
experiment not the Miller experiment that was the exception to this
rule. Models frequently provide useful insight into which results will
be reproduced and which will not. Einstein's faith in special
relativity turned out to be justified.

In some cases naive falsification does work and models are
abandoned. For, example the minimal $SU(5)$ grand unified theory in
particle physics was abandoned because the proton did not decay as
predicted. On the other hand, the quark model of hadron structure was
accepted because of the discovery of the $J/\psi$
particle\cite{jpsi}. The confirmation of a striking prediction gives a
model significant credibility and is the surest route to get a model
accepted. Thus, the discovery of the three degree Kelvin microwave
background led to the acceptance of the big bang model of the
universe. The alternate models were not falsified so much as the new
models, the quark model or big bang model, shown to have superior
predictive power. These models were not developed by scientific
induction, which Hume showed does not really exist, but rather were
tested by observation and survived.
 
Predictive power is related to testability. To be science, models must
make predictions that can be tested against future observations.  It
is the predictions that are falsifiable rather than the models.
Describing past observations (sometimes referred as to postdictions or
retrodictions) is necessary but not sufficient.  It is easy to describe past
observations. For example: Why is the sky blue? Why is the mass of the
proton $1,6726 \times 10^{-27}$ kg?  Why is the mass of the electron
$9.10938188 \times 10^{-31}$ kg?  Answer: Because God (or whoever)
made it that way. While this may be true and can account for all
observation it is completely lacking in predictive power and suggests
no new research directions. It is not science but rather the end of
science.

While we have discussed how models are tested we have not discussed
how models are constructed. These are two very different procedures
both logically and practically. The testing is done through comparison
with observations. But before the testing we need a model to test. The
pre-1905 view was that models were inductively deduced from
experiment. The ``apple down'' approach quoted from
Gopnik\cite{Gopnik}. In this view the creation and testing were
inextricable linked together with the model deduced from the
observations through induction.  This is not correct. Rather model
construction is a creative activity --- as creative as anything in
literature or the fine arts. There is no algorithm saying how to go
from observations to a model.  A falling apple inspired Newton, rising
water in a bath inspired Archimedes, a dream inspired Kekul\'e (the
structure of benzene). Or at least that is how the stories go. For
model construction, Feyerabend\cite{Feyerabend} is correct, anything
goes --- dreams, divine inspiration, pure luck, or even hard
work. General relativity, inspired by the equivalence between inertial
and gravitation mass, was largely due to Einstein's creative genius
and his concept of elegance.  However, it is the testing of models
that separates science from other human endeavors.  Regardless of how
the model is constructed or how elegant it is, to be science, it must
be tested against careful reproducible observations.

\section{To Err is Human, to Control Errors, Science}
\label{sec:err}

The first thing one learns in trying to gain knowledge through
observation is that it is easy to make mistakes. Thus much of the day
to day work of a scientist is error control. In first year university
physics courses you are taught that a measurement consists of the
measured number plus an estimate of the error; the so called error
bar. In science, a measurement without an error estimate is mostly
useless.

Most exciting new results are wrong --- canals on Mars,
Prof.~Blondot's n-rays\cite{Ashmore}, polywater\cite{Franks}, the 17
kev neutrino\cite{Wietfeldt}, cold fusion\cite{Close}, superheavy
element 118\cite{Dalton} and penta-quarks\cite{Hicks}. The list goes
on and on. Occasionally one survives like the $J/\psi$ discovery in
particle physics.  The ones that survive become well known, everyone
quotes Galileo (1564 - 1642), while the ones that are wrong, like
Prof.~Blondot, are forgotten. All physicists know about Michelson and
Morley, very few about Dayton Miller. Thus there is a popular
misconception about scientists being unduly harsh on new ideas and
unexpected results. Science is justifiably leery of accepting new
unexpected results. The degree of fit with existing models is a
valuable, but not decisive, criterion in judging new results.  The
weight of evidence for an extraordinary claim must be proportioned to
its strangeness\cite{Laplace}.  Unlike in the justice system,
unexpected new results are assumed ``guilty'' until ``proven''
innocent. It is up to the proponents of the new result to convince the
rest of the community that the results are correct. It is not up to
the rest of the community to show the new results are wrong. Since
there are many more ways, an uncountable infinity of them, to do an
experiment wrong than to do it correctly, one has to be very critical
of new results, especially unexpected ones. It is very easy to fool
oneself. Prof. Blondot's ``discovery'' of n-rays is an archetypal
example of self delusion.  From the examples given, it might seem that
error control is mainly important for observations. This is not
true. Errors can also occur in the construction and use of
models. Models can be internally inconsistent or their predictions may
be incorrectly calculated.

So how is error controlled? First, observations and model testing must
be done carefully with attention at every step to the possibility of
error.  One of the hallmarks of science as compared to pseudo-science
(see ref.~\citen{pseudo} for a discussion of pseudo or pathological
science) is the concern taken to control errors --- double blind
clinical trials in medicine, blind analysis in physics for
example. The most important discovery of modern medicine is not
vaccines or antibiotics, it is the randomized double-blind test, by
means of which we know what works and what doesn't (quoted from
ref.~\citen{Park}). Strict error control on both observations and models 
is necessary to 
prevent science from being overrun with bogus results. Alternate 
medicine proponents, for example, do not seem to pay enough 
attention to error control.

Secondly, results must be independently reproducible. Different
independent scientist must be able to reproduce the results. To be
reproduced, the results must be made known and made known in
sufficient detail that others can understand how to repeat the
procedure. The experience is that results that can not be repeated are
probably wrong. An unexpected result can be due to error or in some
cases outright fraud. Occasionally it may be correct. Repeatable
observations, even unexpected ones, are probably correct. Independent
replication is a strong line of defense against both error and
fraud. In some cases independent replication is difficult or nearly
impossible. For example, high energy physics experiments are very
expensive and even nominally independent experiments share common
features like the inputs to Monte Carlo computer simulations. In this
case even greater care must be taken. In medicine, there is the
additional problem of the trade off between additional tests and not
immediately implementing a new treatment that {\em may} save lives.

Repetition is not doing exactly the same experiment again and
again. Rather, the subsequent experiments should be as different as
possible to eliminate common sources of error. The first experiment is
usually the most difficult. It identifies an interesting effect and
says where to look to find it again. The second experiment can then
use this information to refine the technique. Consider again the
Michelson-Morley interferometry experiment. The following summary is
taken from ref.~\citen{Shan}. The initial experiment was done in 1881
by Michelson. The limit on the shift in the interference line he
obtained was half that expected based on the ether model. The famous
Michelson-Morley experiment in 1887 had reduced that to 0.025 of the
predicted shift. An experiment in 1930 by Joos reduced that to 0.0027
of the predicted shift. Dayton Miller in 1926 was claiming an effect
of 0.077, much larger than Joos but still much less than expected from
a simple application of the ether model. Ignoring Miller for the
moment, we have over fifty years an improvement of over two order of
magnitude. Modern experiments with lasers have reduced this an
additional sixteen orders of magnitude\cite{Brillet}. Thus we see the
initial experiment repeated over time, with different techniques, and
dramatically improved. The Miller result is an anomaly but not an
unique one. In any sequence of measurements it is usual for one or two
to be wrong by more than expected from the quoted error. In difficult
experiments it is easy to make mistakes. The subsequent, much more
accurate measurements, make the Miller results only an historical
curiosity. It is not necessary to explain exactly what the error
was. However, ref.~\citen{Shan} does suggests thermal gradients are at
least partly to blame. Again we stress the role of repeatability to
establish experimental or indeed any results. In science, as this
example illustrates, the observations become more accurate over time
as the experiments improve. A sure sign of pathological science or
pseudo-science is when the signal does not improve but stays at the
barely detectable limit as the experiment is improved.

Reproducibility does not mean science cannot study historical events
because the event cannot be reproduced. When a plane crashes, it is
possible to use science to learn what caused the crash without waiting
for future crashes. Rather one models the crash and tests predictions
of the model against observations of the properties of similar the
planes and the materials making them up and also against observations
of the debris from the crash. Other historical events can be studied
similarly. In the case of evolution one studies modern biological and
ecological systems and the ``debris'' in the form of fossils and the
present distribution of species. Thus evolution is tested by comparing
its predictions against observations of these.

Next for error control is peer review. This is a simple concept. The
people who know about a topic are peers of the person who did the
original experiment and they look to see if there are any errors. It
is only the peers who would have the knowledge to spot errors. The
first line of peer review is the informal discussions a scientists has
with his colleagues. Many errors are caught at this stage. The second
line of peer review comes from the anonymous reviewers who act as gate
keepers for scientific journals. Third is the review a paper receives
after it has been published. The first and third types of peer review
are much more important than the second. Important papers are probably
correct, not because the author is infallible, but because many people
have independently checked the results and looked very hard for
errors. It is only after this independent checking that a paper should
be assumed correct. Unimportant or less studied papers probably lack
the third level of review and should be treated more suspiciously. The
third type of peer review can take place without the second or even
the first. This happens to papers that appear on the web archives
(lanl.arxiv.org for example) with critical papers commenting on them
appearing before the first paper is formally published. This is quite
valid and very important peer review.

Secrecy and intellectual property rights when they lead to secrecy are
the enemies of scientific progress since science depends on the
sharing and checking of results for error control. Two heads are
definitely better than one. Without the independent checking, errors
last for a long time --- especially when there is an economic gain
from keeping the error hidden.

The peer review process also works in software development. The Free
(free as in speech not beer) Software and the Open Software Movements
encourage a peer review process for software development. Program
source code is made widely available usually on the world wide web.
People can then build on the work of others. It was discovered that
many eyes looking at programs found and fixed errors
faster\cite{Raymond} than in other approaches. The internet protocol,
the world wide web, the Apache web server (which powers the majority
of web sites in the internet), the Linux kernel and the Fedora
operating system, to name a few, are the products of an open
development process. This illustrates the power of openness and peer
review that is the hallmark of science.  ``Given enough eyeballs, all
bugs are shallow''\cite{Raymond} applies equally to science and
software development.

\section{The Graveyard of Departed Models}
\label{sec:gdt}

\subsection{The Relativity of Wrong}

What happens to models that have been disproven? Are they buried
somewhere with a large tombstone? No, either they are forgotten or,
like the Night of Living Dead, they stay around forever as
approximations to more complete models. The latter idea is lucidly
presented in an article by Isaac Asimov called ``The Relativity of
Wrong''\cite{Asimov}. The ideas expressed there are as important as
Popper's, Kuhn's or even Hume's in understanding science. The basic
concept is that one should not ask if a model is right or wrong but
rather how wrong is it. Experimental error bars are the
experimentalists' estimate of how wrong the experimental number might
be.

To clarify the relativity of wrong concept, consider the value of
$\pi$. A simple approximation is $\pi=3$ (1 Kings 7:23) . This is
wrong but by less than 5\%. A better approximation is $\pi=3.14$. The
error here is 0.05\%. Strictly speaking both values are wrong. However, the
second value is less wrong than the first. In computing as a graduate
student I used $\pi=3.141592653589793$. This is still wrong but much
less wrong than the previous approximations.  There was no sense using
a more accurate value of $\pi$ since the computer used only about 15
digits (single precision on a CDC computer).  We see that wrong is a
relative concept. None of the values of $\pi$ are absolutely
correct. That would take an infinite number of digits, so all are
wrong.  However the initial values are more wrong than the latter
values. They all are useful in the appropriate context. The value from
Kings probably reflects the accuracy with which the measurements were
made.

The same logic applies to models. Consider the flat earth model (see
also the discussion in ref.~\citen{Asimov}). For the person who never
travels farther that 100km from his birthplace the flat earth model is
quite accurate. The curvature of the earth is too small to be
detected. However when the person is a sailor the question of the
shape of the earth takes on more urgency. The flat earth model
suggests questions like: Where is the edge of the earth?  What will
happen if I get too close?  For the world traveler, the flat earth
model is not sufficient. The spherical earth model is more useful, has
greater predictive power and suggests a wider range of
questions. Questions like: Does the earth rotate? Does it move around
the sun or does the sun move around the earth? But it is a wrong
statement that the earth is exactly spherical. Not as wrong as the
statement the earth is flat but still wrong. However being not exactly
correct does not make it useless. A spherical globe allows a much
better understanding of airplane routes than a flat map. But the earth
is not a perfect sphere. It is flattened at the poles (a quadrupole
deformation). Smaller still is its octapole deformation. The exact
shape of earth will never be measured as that would require, like
$\pi$, an infinite number of digits. It would also be useless. What is
needed is a description sufficiently accurate for the purpose it is
being used for. Science is the art of the appropriate
approximation. While the flat earth model is usually spoken of with
derision it is still widely used. Flat maps, either in atlases or road
maps, use the flat earth model as an approximation to the more
complicated shape.

Classical mechanics --- Newton's law of motion and Maxwell equations
of electromagnetism --- although superseded by relativity and quantum
mechanics are still useful and taught. The motion of the earth around
the sun is still given by Newton's laws and classical optics still
works. However, quantum mechanics has a much wider realm of
reliability. It can describe the properties of the atom and the atomic
nucleus where classical mechanics fails completely.

Animals reproducing after their kind is the few generations limit of
evolution. Thus over the time scale of few human generations we do not
see new kinds arising. The offspring resemble their parents. Evolution
keeps the successes of the previous model; cats do not give birth to
dogs nor monkeys to people even in evolution.  The continuity between
animals-reproducing-after-their-kind and evolution is not sufficiently
appreciated by the foes of evolution and perhaps not by its proponents
either.

The Ptolemaic model with the earth at center of the universe can be
considered as an approximation to the Copernican model. In fact a very
useful one. When giving directions to the corner store, an earth fixed
model is used. The directions in an heliocentric model would be quite
complex and ghastly to contemplate. Actually this case is even more
subtle. According to the general theory of relativity, the laws of
motion can be expressed in any inertial or accelerated frame. Thus the
choice between a heliocentric model and an earth-centric one is not a
matter of right or wrong but one of convention and convenience. What
is a model assumption and what is convention is not always clear.
Poincar\'e, one of the leading mathematical physicists of the late
19th century, claimed\cite{Poincare} that much of what is regarded as
fact is convention.

There is a general trend: new models reduce to the previous model for
a restricted range of observations. Ideally the new model would
contain all the successes of the old model, the correspondence
principle of quantum mechanics applied more widely, but this is not
always the case. A subset of the correct predictions of the previous
model may not survive. This loss of predictive power is sometimes
called Kuhn loss. But overall the new model must have more predictive
power otherwise its adoption is a mistake. Thus we have the view of
science producing a successions of models, each less wrong (none are
100\% correct) than the one it replaces and we see progress. Science
does progress, new models are constructed with greater and greater
predictive power. The ultimate aim is to have a model of everything
with a strictly limited number of assumptions. This model would
describe or predict all possible observations. Quantum indeterminacy
suggests that such a model does not exist. However, progress in
science is moving closer to this ultimate probably-unreachable goal.

It is worth considering further science as the art of the appropriate
approximation. Exact calculations are never done. Consider calculating
what happens when you drop a pencil. You start with Newton's law of
gravity and Newton's laws of motion. This is a very good
approximation. Next you have to consider the effect of the interaction
with the air. This may be important on very windy days or for long
drops. Then there are the effects of special relativity, general
relativity and quantum mechanics. There are also the tidal forces due
to the moon, the sun, Mercury, Venus, Mars, Jupiter, Saturn, Neptune, Uranus,
Pluto and the asteroids. Even including all this is not
sufficient. There is also the gravitational effect of the stars, the
cat and the canary. This is clearly ridiculous but so is the idea of
an exact calculation. So we do not do an exact calculation. Rather we
include only the effects that are large enough to have a noticeable
effect on the pencil. Classical mechanics and Newtonian gravity are
probably enough although in some cases the air may have an observable
effect (the terminal velocity of an object dropped from a great height
is due to air resistance). One of the most important things in science
is deciding which effects are large enough to have to be taken into
account and which can be neglected.

\subsection{Paradigm Shifts}

Kuhn discusses two kinds of science --- normal science and
extraordinary science. Normal science is puzzle solving within the
context of a paradigm while extraordinary science is the overthrowing
of the paradigm. The present analysis gives a different view of the
distinction although the distinction itself remains useful. In
extraordinary science the model being challenged is the main model in
the field, the paradigm (or model) which provides the framework for
the field. In normal science it is the subsidiary models, those models
that in principle could be derived from the main model, that are being
tested. Frequently it is only in retrospect that it becomes clear
which type of model is under attack. Normal science can often resemble
scientific induction with models apparently deduced from
observations. This is especially true in cases like the color of crows
or Hooke's Law where a regularity is apparent in the observations.
Operationally, scientific induction can be considered an approximation
to the more general method of model building and testing. However, the
approximation is only useful in a limited range of situations. General
relativity or even the structure of benzene can not be considered to
have been derived by induction.

When there is a paradigm shift, {\em i.e.} when the main model in an
area of research changes, the new model usually expands the range of
observations that can be described. Quantum mechanics describes both
microscopic and macroscopic objects. Maxwell's equations described
electricity and magnetism with one model. Special relativity plus
quantum mechanics (quantum field theory) describes both mechanics,
like Newton's law and electromagnetism like Maxwell's equations but
over a wider range of energies. While the reproducing-after-their-kind
model applies mainly to plant and animal husbandry, evolution also
describes fossils and the distribution of species. Basically the old
model worked well for a limited range of observations. Paradigm shifts
typically occur when the old model is being pushed into a new area
that has not been explored before or not explored in as much
detail. The new model must describe both the observations in the
regions where the old model worked and in the new regions where it
doesn't.

The picture, just given, is of an orderly progression striving to
ultimate perfection. But something else is happening, especially with
extraordinary science. Consider the case of Newtonian mechanics
replacing Aristotelian mechanics. While a vase slid across the table
still came to a stop, hopefully before reaching the edge, the
Aristotelian concept of causality broke. Much of the concept of cause
and effect was forever altered. Newtonian mechanics, in turn, led to
the concept of the clock-work universe. With the advent of quantum
mechanics the clock-work analogy no longer applied. The basic
understanding of reality changed. The basic understanding of how
science progressed also changed. With the advent of special relativity
the ether disappeared. From the present view point this looks like a
small change but consider the definition of physics given in a 1902
high school physics text book\cite{Gage}: ``Physics is the science
which treats of matter and its motion, and of vibrations in the
ether''. Thus the very understanding of the nature of physics changed.
The kinetic theory of heat destroyed caloric but the Carnot cycle and
Carnot's principle developed on the basis of the caloric model
survived.  Evolution changed the basic understanding of biology and
the relationships among different species including {\em Homo
sapiens}. Its theological repercussions are still vibrating. The
discovery of non-Euclidean geometries and their application in special
and general relativity destroyed the Kantian idea that Euclidean
geometry is synthetic {\em a priori} knowledge and perhaps even the
idea that synthetic {\em a priori} knowledge is possible.

This change of world view when the main model in a given field of
study changes is closely related to Kuhn's
incommensurability\cite{Kuhn} (see also
Feyerabend\cite{Feyerabend}). Proponents of the new model and the old
model use a different language and different concepts. They have
different ideas about what the important questions are. Their whole
framework for understanding observations is different. Hence, it may
be difficult to compare the old and new models in detail.  Despite
these dramatic differences, the Ptolemaic system, the Copernican
system, Newtonian mechanics, special relativity, quantum mechanics and
general relativity can all quite accurately describe Jupiter's
apparent motion through the night sky. If string theory wants to be
accepted as the model of everything it too must describe Jupiter's
motion.  Thus we have, on the one hand, the previous models continuing
to be good approximations to the new models at least for a limited
range of observations while, on the other hand, the philosophical and
metaphysical implications having a profound and frequently disturbing
break. When the main model in an area changes the perception of
reality can change in dramatic and unexpected ways. Philosophy and
metaphysics based on the current models of science are very precarious
and can be rendered obsolete by new and improved models. If the past
is any guide, the next great advance in science will change our
understanding of the nature of reality and/or the relation of humans
to the rest of creation. However, animals will still reproduce after
their kind as stated in the book of Genesis and the earth's motion
around the sun will still be accurately described by the model Newton
developed in the 17th century. Old models live on as approximations to
later models. However, the metaphysics and internal constructs of old
models are buried without headstones where only historians can or
would want to find them.

\section{Scientific Equivalence}
\label{sec:se}

In science, models are judged on their ability to describe and predict
observations. If two models give the same results for all observations
they are scientifically equivalent. In quantum mechanics there is a
mathematical technique known as unitary transformations that while
making the mathematics quite different leave all predictions of
observables the same. Canonical transformations play a similar role in
classical mechanics. Observations can not distinguish between such
models and it can be argued that equivalent models are not really
different.

There are more low brow examples of equivalent models. Consider Last
Thurs\-day\-ism\cite{Thursday}. This is the model that the universe
was created last Thursday but in such a way that it is
indistinguishable from an old universe.  If we had the universe
created 6000 years ago rather than last Thursday this is essentially
the Omphalos Hypothesis\cite{Gosse}, a 19th century attempt to
reconcile geology with the a literal interpretation of the Bible. The
light from distant stars, which would not be seen if they just started
shining last Thursday, is created in transit. Memories are created in
place without referring to actual events. Similarly everything else is
created to be indistinguishable from an old universe. Since by
construction this model has all the same predictions as the old
universe model the two models can not be separated based on
observations. The only criteria is simplicity. Last Thursdayism is
rejected because it has an additional assumption, namely that the
world was created last Thursday.  This assumption does not increase
the model's predictive power.  While this example is contrived, it is
not trivial and does illustrate the point.  It is easy to construct
equivalent models with rather different content with only simplicity
(or prejudice) to use to eliminate them.  This use of simplicity is
sometimes called Occam's razor after William of Occam who said ``one
should not increase, beyond what is necessary, the number of entities
required to explain anything''.  A modern paraphrase is that there
should be no more assumptions than the minimum needed. Since it easy
to change the internal properties of a model without changing the
predictions great care should taken in attaching meaning to such
features of the model.

In addition to full equivalence there is effective equivalence. For
example, quantum mechanics and classical mechanics give the same
result for planetary orbits to the accuracy of any foreseeable
calculation. Thus for planetary orbits the two models are effectively
equivalent.  For objects moving much less than the speed of light
special relativity and classical mechanics are effectively
equivalent. For a low number of generations evolution is effectively
equivalent to animals reproducing after their kind. More generally
when one model provides a good approximation to another model for some
range of observation they can be considered effectively equivalent for
that range of observations.

The philosophical idea of realism is that behind observations there is
an objective reality. The observations are the same whether or not
there is an objective reality. Thus models with and without objective
reality are scientifically equivalent. The only tool in our tool box
to deal with this situation is Occam's razor or equivalently
simplicity. {\em A priori} it might seem that the existence of
objective reality is just an additional assumption that simplicity
says we should eliminate. However, objective reality is an integral
part of most scientific models. For example, in both the Ptolemaic and
Copernican models, the earth and planets are objectively
real. Removing the objective reality from these models would involve
an additional and unnecessary assumption. This is much like Last
Thursdayism where everything since last Thursday is assumed to have
objective reality but things before last Thursday are assumed not to
have objective reality. By moving the cutoff time from last Thursday
to the present time we have essentially removed all objective
reality. One could also have Next Thursdayism where things only take
on an objective reality after next Thursday. Models without objective
reality can usually be eliminated, like Last Thursdayism was
eliminated, by the use of Occam's razor. Keeping realism as an
integral and significant part of the models while maintaining the
models tentativeness leads to critical realism.

Another example of equivalence is the many worlds
interpretation\cite{Everett} of quantum mechanics. The fundamental
idea of this interpretation is that there are myriads of worlds in the
universe in addition to the world we are aware of. In particular,
every time a quantum experiment with different possible outcomes is
performed, all outcomes are obtained, each in a different
world. However, we are aware only of the world with the outcome we
have seen. To the extent this is just an interpretation it is
scientifically equivalent to the usual Copenhagen interpretation {\em
i.e.}  all predictions will be the same. There is an argument about if
the two are practically equivalent rather than fully equivalent but
for the present discussion this is not significant. The many-worlds
interpretation is just an extra assumption that has no testable
consequences. It can, therefore, be eliminated by Occam's razor.

One could also have a theistic interpretation of quantum mechanics in
which God determines exactly which outcome will occur but lets mere
mortals only predict the probability. Thus, although God does not play
dice (perhaps keeping Einstein happy), to mortals, through ignorance,
it appears he does. It also allows God to control the evolution of the
universe without violating any physical laws. The theistic
interpretation is equivalent to ordinary quantum mechanics and just
provides another way around the philosophical implications of quantum
mechanics that some people find repugnant. Like the many worlds
interpretation it can be eliminated from science by the use of Occam's
razor.

The present work suggests, rather than a many worlds or theistic
interpretation, an instrumental or phenomenal approach to quantum
mechanics. The important and lasting part of quantum mechanics is the
mathematical formulation which provides the basis for making
predictions for observations. As the theistic interpretation
illustrates, different conflicting interpretations can be easily
constructed. Poincar\'e expressed the role of mathematics even more
strongly\cite{poincare} ``But what we call objective reality \ldots
can only be the harmony expressed by mathematical laws. It is this
harmony then which is the sole objective reality, the only truth we
can obtain.''

From the point of view of predicting and describing observations the
statements ``God made it that way'' and ``It is that way because it is
that way'' are equivalent. Both perfectly describe all past
observations and predict no future observations. Each observation
explained this way needs its own assumption. The cosmic
anthropological (or anthropic) principle is also equivalent. The
cosmic anthropological principle states that the reason the universe
is the way it is, is because otherwise people would not exist.
However we could equally have a zoological principle, or an
agrostological principle, or a geological principle or a planetary
principle. These principles all are just the statement that the
universe is the way it is because if it wasn't it would be
different. The different part being emphasized by the anthropological
principle is the presence or lack of people.  The agrostological
(agrostic) principle, in contrast, emphasizes grasses: farmers are
grasses' way of competing with trees and golf courses mainly benefit
grass. The ``It is the way it is because that is the way it is'' or
equivalently ``God made it that way'' are show stoppers in
science. They have no predictive power and are essentially an
admission of defeat. It cuts off further progress in the given
direction.

There is another way to construe the admission of defeat. Namely, some
of the parameters in the model have to be determined
phenomenologically. In the past there has been a need for
phenomenological input, but this always was assumed to be due to
ignorance or uninteresting initial conditions. The mass of the proton
is determined phenomenologically but could be determined, in principle
from Quantum Chromodynamics. The mass of the earth is determined
phenomenologically and is presumably due to conditions in the early
solar system. Now we have the possibility that more fundamental
parameters like the mass and charge of the election must be determined
phenomenologically.  The situation is rather like statistics in
quantum mechanics. Before quantum mechanics, statistical approaches
were used but only as approximations to deterministic models. Quantum
mechanics, in contrast, is inherently statistical.  If the idea being
discussed in this paragraph is correct, parameters that were
previously believed to be obtainable theoretically will only be
determinable phenomenologically. Ideally all the parameters in a model
would be determined theoretically. This is probably too much to hope
for and would reinstate Kant's synthetic {\em a priori}
information\cite{Kant} --- nontrivial information determined without
phenomenological input.
 
\section{Science and Religion}

\subsection{Observation {\em vs.}\ Divine Revelation}

Science and religion are not always in harmony, whether Galileo versus
the Catholic Church or the modern anti-evolution teaching of some
churches. It is therefore important to understand the relation between
the two.  It is not fundamentally one of faith versus logic. Any
system of epistemology has assumptions, even science. While religion
places more emphasis on faith, many claim their religious beliefs are
based on sufficient reason. Wright in a series of books
(ref.~\citen{Wright} and other books in that series) argues for the
historicity of Jesus from historical analysis. Both science and
religion use models. The literal interpretation of Genesis is a model
for the creation of the world. So the difference is not in model
building versus absolute truth.

Similarly the distinction based on natural versus supernatural is not
valid. At one level the introduction of the supernatural in the form
``Because God made it that way'', as discussed above, is a show
stopper in science. The problem is not that it is supernatural but
that it lacks predictive power. At another level saying that science
is based on natural explanations is just a tautology since
supernatural can be defined as that which has no explanation in the
current models of science. The National Academy of Sciences
says\cite{science} ``Anything that can be observed or measured is
amenable to scientific investigation.''  Presumably this is true even
if the thing being observed is considered to be supernatural. For
example Matthew 17:20 says\cite{NASB} ``for truly I say to you, if you
have faith the size of a mustard seed, you will say to this mountain,
`Move from here to there,' and it will move; and nothing will be
impossible to you.''  Clearly mountains moving around would be
observable and thus amenable to scientific study. Hence the scientific
method is not methodological naturalism\cite{Vries}. Merely because
something is supernatural does not exclude it from scientific
study. The question of supernatural phenomena in situations like this
should be settled by observation, not {\em a priori}. It is amazing
how far observation-based models have gone in describing observations
without explicitly invoking the supernatural. To some extent this is
because once science describes something it is no longer considered
supernatural. In part, scientists' reluctance to admit the explicitly
supernatural into their work is because so much success has been
obtained without it. It is also in part because the supernatural
usually adds no predictive power. While I know of no studies of the
effect of prayer on mountains, there have been controlled studies on
the effect of prayer on healing\cite{Benson} showing little or no
effect. Presumably the biblical quote above is not meant literally
otherwise most Christians must have faith smaller than a mustard seed.

So what is the distinction between science and religion?  The main
difference is that in science the ultimate authority is careful
observation with an emphasis on predictions while in religion it is
divine inspiration or revelation. The inspiration may come directly
through a person's religious experiences in their daily lives or the
revelation may come through a prophet or sacred texts. The Vedas, the
Torah, the Christian Bible, the Koran, and the Guru Granth Sahib are
examples of such scared texts. When the divine revelation comes
through a person that person becomes the authority. The difference in
the two approaches is highlighted in the conflict between Galileo and
the Catholic Church. Galileo looked through his telescope while the
church leaders consulted the Bible. One can debate whether the church
officials correctly interpreted the Bible but one can not dispute that
it was their ultimate authority. They did not look through the
telescope since they ``knew'' the scriptures were more important than
observations.

What Galileo was proposing was very revolutionary. That the earth goes
around the sun was the least revolutionary part. The real
revolutionary part was that you should look through the telescope and
compare what you saw with what the church and other authorities were
teaching. This was a whole new way of studying reality. It also
directly questioned the authority of the church. The church leaders
naturally reacted with horror. Their whole edifice of belief and
authority was under attack. They accused Galileo of undermining
Christianity. Eventually an uneasy truce was reached between science
and religion.  Observation would be used to study natural history. The
relationship between God and man, and morality would be left to the
church. The truce was broken by evolution which many in the church
regarded as challenging their understanding of the relationship
between God and man. It was considered a threat to their religion, a
parallel to the reaction to Galileo. Similar skirmishes are expected
in the future as observation-based science invades more of the
traditional domain of religion.

In principle, observation based models and models based on divine
revelation do not have to be in conflict. There is no reason a model
created through divine revelation could not have predictive power ---
in creating models anything goes. The conflict arises when a model
based on someone's divine revelation or interpretation of divine
revelation conflicts with models based on observations. In this case
one has to decide between observation and divine revelation. Atheists
and deists would claim divine revelation does not even exist. The
defense of divine revelation and the thorny problem of deciding
between the different purported divine revelations is not part of the
present work but rather is left to the theologians.

One of the traditional methods to reconcile science and religion is
the God of the gaps theology\cite{Drummond}. This says that science
accounts for what it accounts for and whatever is leftover (the gaps)
is due to God. The cause of the big bang and the origin of life are
two of the gaps in current scientific understanding. A miracle
defined\cite{AHD} as ``An event that appears inexplicable by the laws
of nature and so is held to be supernatural in origin or an act of
God'' is another example of the gap concept.  Irreducible complexity
has been advanced\cite{Behe} as an attack on evolution and is, along
with much of creationism, very much in the God of the gaps
tradition. The God of the gaps theology attributes everything not
currently described by scientific models as due to God either directly
or by implication. This is a special case of the more general
technique known as ``Proof by lack of imagination''. The argument goes
like this: I can not imagine how this can happen naturally therefore
it does not or God must have done it. This argument only works until
someone with more imagination comes along. For the example of
irreducible complexity, the judgment\cite{Dover} in the Kitzmiller
{\em vs.}\ Dover Area School District court case found that many more
imaginative people have already come along. The general problem with
the God of the gaps theology is that as more imaginative people come
along, the gaps disappear and so does the God. With the present view
of science as being primarily descriptive, the God of the gaps
theology makes little sense anyway. God, if he/she/it exists, would
presumably be responsible for what science describes as well as for
what it does not describe. To the deist or theist, science is simply
the description of God's handiwork as made manifest through
observations.

The crux of the science {\em vs.}\ religion debate is where to
demarcate the boundary between where we allow observation to be the
final authority and where we allow sacred texts or sacred texts
interpreted by religious leaders to be the final authority. There are
advocates of both extremes and many places in the middle.  Deuteronomy
18:22 suggests that if the ``divine revelation'' does not come true
{\em i.e.} is not consistent with observations ``that is a thing that
the LORD has not spoken''\cite{NASB}. Hence the worry on the part of
some Christians that if the Bible is not literally true the LORD has
not spoken it.  Be that as it may, the semantic argument is very
clear: models, beliefs and natural history based on divine revelation
are religion, models and natural history based on careful observations
are science, models and beliefs based on less than careful
observations are pseudo-science and superstition.

The difference between science and religion is that science has
observation as its ultimate authority while religion has divine
revelation. Because of this basic difference other differences
follow. Since science is based on observation it appears
materialistic. Religion being based on divine revelation assumes that
the divine exists and is important. The role of individual people is
also fundamentally different. Christ is central to Christianity: 1
Corinthians 15:14\cite{NASB} ``and if Christ has not been raised, then
our preaching is vain, your faith also is vain.'' Islam is based on
the idea ``There is no God but Allah and Mohammad is his
prophet''. Individuals also play a leading role in defining other
religions and philosophic traditions --- Abraham and Moses (Judaism),
Buddha (Buddhism), Confucius (Confucianism), Lao Tzu (Taoism), Guru
Nanak (Sikhism), Zoroaster (Zoroastrianism), and Bah\'a'u'll\'ah
(Bah\'a'\'i Faith).  Even at an operational level certain people have
an elevated position and are considered authorities, for example the
Pope in the Catholic Church or the Grand Ayatollahs in Shiite
Islam. In science, on the other hand, the people are incidental. If
Charles Darwin had not been there, Alfred Russel Wallace would have
been known as the discoverer of evolution. At the present time,
neither Darwin nor Wallace are relevant to the validity of
evolution. The validity of relativity does not depend on Einstein's
greatness nor quantum mechanics on Bohr's. Rather Darwin, Einstein and
Bohr are considered great because of the greatness of the models they
helped develop. Darwin, Einstein and Bohr are not the ultimate
authorities in science, the observations are. When people are taken as
the authority in science the field stagnates. An example is the
physics in England after Newton. The physicists were so in awe of
Newton that they used him as the authority and thus fell behind their
colleagues on the continent. When anti-evolutionists attach Darwin to
modern evolutionary models and try to discredit them through him,
scientists look on in bewilderment. Conversely, the western largely
secular and observation based society does not understand why the
Muslim world is so upset at attacks or perceived attacks on
Mohammad. An attack on Mohammad is an attack on the center of the
religion he started. It calls the whole of the religion into question
since Mohammad is {\em the} authority that defines Islam. A similar
argument holds for Christ and Christianity.  On the other hand, an
attack on Darwin, Einstein or Bohr is an irrelevancy.

Finally we note that having science and religion cooperate is not
always a good thing. In warfare, the motivation is all too frequently
provided by religion and the means by science.

\subsection{Creationist Natural History}

There is currently an attack on evolution by one segment of the
Christian Church.  This comes across as an attack on Darwin and what
he represents. However their argument is really with Galileo and his
telescope.  Once we are allowed to take what we see through the
telescope, {\em i.e.} observations, as the ultimate authority rather
than scripture the rest of science follows.  See for example ``The
Wedge Strategy''\cite{Wedge} where the attack on evolution is promoted
as the beginning of an assault on natural history based on observation
(materialism) rather than scripture. It is the thin edge of the wedge
since Darwin is considered more vulnerable than Galileo. They want
to\cite{Wedge}: ``To replace materialistic explanations with the
theistic understanding that nature and human beings are created by
God.'' In principle, there is nothing wrong with developing a natural
history based on theistic concepts derived from divine revelation.
However, enlightenment thinkers like Kant and Voltaire argued that the
Middle Ages were the Dark Ages because independent thinking and
observation took a back seat to divine revelation or, at least, the
church leaders' interpretation of divine revelation. In any case,
natural history based on divine revelation rather than observation is
religion not science. Note the distinction between natural history and
science. Natural history is any study of how nature works. To be
science it must be based on careful observations. Creation science and
creationism are not science since the ultimate authority in this field
of study is the Bible rather than observations. They are religion and
creation science should rather be called creationist natural
history. How accurately it represents Biblical authority is another
issue and is again left to the theologians.

The evolution-creationism controversy provides insight into how
science works.  Evolution is attacked as being unreasonable. As
previously noted, many successful models have been rightly considered
unreasonable.  In addition, unreasonableness, like beauty, is in the
eye of the beholder. Many scientists, myself included, regard natural
selection as an extremely intelligent way to do design,
resembling\cite{IEEE} Monte Carlo techniques in computer
science. Another line of attack on evolution is through ``proof by
lack of imagination'' as typified by irreducible complexity. Proof by
lack of imagination is related to naive falsification: I can not
imagine how your model can describe this result hence your model must
be wrong.  However, with enough imagination any negative result can be
explained away. The white crow is really a black crow covered with
snow or the sun is reflecting (specular reflection) off the black crow
in such a way it looks white.  Thus attempts to attack evolution
through naive falsification will fail. Similarly attacks on
creationist natural history based on naive falsification will
fail. Although the Omphalos hypothesis is {\em ad hoc}, like Last
Thursdayism, it can only be eliminated by appealing to
simplicity. Imaginative people will find ways around any possible
falsification.  However, modifying models to circumvent falsification
usually reduces their predictive power. In the case of creationist
natural history and the Omphalos hypothesis it destroys it altogether.
As the 1902 textbook quote\cite{Gage} illustrates, the 1887
Michelson-Morley experiment did not immediately falsify the ether
model. Creative people came up with explanations such as ether
entrainment and the Lorentz-Fitzgerald contraction to explain the
unexpected results.  Both explanations reduced the ether's predictive
power.  In the end, the ether model was eliminated by Einstein's
special theory of relativity which had fewer assumption and more
predictive power than the competing models.  Michelson and Morley
provided the ammunition, Lorentz the gun (the Lorentz transformation),
but it was Einstein that pulled the trigger. His defense lawyer would
argue that the ether was not shot but rather had it throat slit with
Occam's razor. Einstein did not prove that the ether did not
exist. Rather he showed that the ether hypothesis, like the Omphalos
hypothesis, has no predictive power, and in the end it was eliminated
by appeals to simplicity.

The proponents of intelligent design argue that there must be an
intelligent agent behind the design of the universe. Unfortunately
intelligent design, like creationism in general, is not well
defined. At one extreme it is considered a replacement for not just
natural selection but even evolution in its entirety. At the other
extreme it is just the theistic interpretation of quantum mechanics
applied more widely; God controlling the universe without explicitly
appearing in the models. The main arguments for intelligent design are
by analogy and proof by lack of imagination. I can not imagine how the
universe could have arisen without a designer, therefore it
didn't. The argument from analogy is based on perceived similarities
between humanly designed machines and biological systems. The
weaknesses in this analogy are given in the Dover court case
decision\cite{Dover}.  The arguments for intelligent design are
remarkably similar to those for the ether: analogy and proof by lack
of imagination. At the time Maxwell discovered the wave equation for
electromagnet radiation, many types of waves were known: sound in
various media, water waves, vibrations in membranes, etc. All had one
thing in common, a physical medium.  By analogy it was argued that the
electromagnetic waves must also have a physical medium --- the ether.
The physicists of the time could not imagine a wave without such a
physical medium, much like some people currently cannot imagine the
lack of a designer for biological systems.  The argument against
intelligent design is the same one that killed both the ether and Last
Thursdayism --- simplicity. Intelligent design, like the ether or the
theistic interpretation of quantum mechanics, is just an extra
assumption that does not lead to new predictions.

The real criteria for judging models is predictive power --- not naive
falsification nor analogy nor lack of imagination. Recently\cite{Dae}
a new fossil, {\em Tiktaalik roseae} --- a tetrapod-like fish or a
fish-like tetrapod, has been found. One can engage in futile semantic
arguments about whether it is a fish, or a tetrapod, or a missing link
or whether it is the work of the devil.  However, the significant
point is that a striking prediction has been confirmed by a
peer-reviewed observation. Using evolution, a model of fossil
formation and a model of the earth's geology, a prediction was made
that a certain type of fossil would be found in a certain type of
rock.  {\em Tiktaalik roseae} dramatically fulfilled that prediction
and provides information on the fish-tetrapod transition. It is just
one of the many strikingly successful predictions of evolutionary
models. Creationist natural history and intelligent design will only
be taken seriously by the scientific community when they are used to
make similarly striking predictions (not postdictions) that are
confirmed by careful, reproducible, peer reviewed
observations. Currently creationist natural history and intelligent
design have only demonstrated minimal predictive power. Evolution
will only be replaced as the dominate model in biology when a
competing model is shown to have more predictive power. Even then
evolution will probably continue as a useful approximation to the more
complete model much like classical mechanics continues to be used as
an approximation to quantum mechanics.

\section{Conclusions}

\subsection{A Meta-Model of Science}

In this paper a meta-model of how science progresses has been
presented. Like a scientific model a meta-model of science must be
logically consistent, as simple as possible and consistent with
observations on how science has developed. As in science, there has
been a sequence of meta-models for how science develops ---Plato's
idealism, Bacon's induction, Cartesian skepticism, Hume's empiricism,
Kant's idealism, Kuhn's paradigms, Popper's falsification and Asimov's
relativity of wrong.  Looking at this progression we see concepts
being introduced and refined --- Kant's ``Critique of Pure Reason'' is
a response to Hume.  Echos of Kant and even Plato are seen in Kuhn's
work. The ``relativity of wrong'' principle applies equally to the
meta-models.  Newer meta-models are less wrong than the previous
models but the older meta-models are not so much wrong as incomplete.
As in science, new observations led to new understanding --- the fall
of classical mechanics played a large role in the motivation and
development of Kuhn's and Popper's ideas.  In addition to emphasizing
the role of the paradigm, Kuhn\cite{Kuhn} also showcased the need to
look at how science has actually developed rather than trying to
develop a meta-model of science based on pure thought. Following his
lead, examples of how science has actually developed are presented in
this work.

The meta-models of science should also make predictions on how science
will develop in the future.  The main predictions of the present work
are: 1) New models will replace older models when and only when they
have more predictive power. 2) The replacement rate will be highest in
fields with highest number of new distinct observations, 3) The
replaced models will be good approximations, {\em i.e.} they will be
effectively equivalent, to the newer models for a limited range of
observations. 4) The philosophical and metaphysical implications of
the new models will be significantly different from that for the
replaced models.

\subsection{Summary}

The scientific method is observationally constrained model building,
not induction, falsification nor methodological naturalism.  The
observations must be carefully done and reproducible. The models must
be logical, internally consistent, predictive and as simple as
possible. Both observations and models should be peer reviewed for
error control. The goal of science is to construct models that make
the maximum number of correct postdictions and predictions with the
minimum number of assumptions. Supernatural explanations are rejected
not {\em a priori} but when, as is usually the case, they lead to no
testable predictions for future observations. In general, if you want
your model to be accepted you must show that it makes more correct
precise predictions with fewer assumptions than the competing
models. There is a surfeit of models that make fewer predictions. As
models are improved their predictive powers increase.  We see progress
with time, the models become less wrong, probably not absolutely
right, but less wrong.  There appears to be convergence towards the
probably unreachable goal of a model of everything. The same can not
be said for the philosophical and metaphysical implications of the
models. Here there is no obvious convergence or at least the
convergence is much slower.  There is no overwhelming reason to
believe the philosophical and metaphysical implications of presently
accepted models. They will probably change in unpredictable ways when
new improved models comes along. The only important, enduring property
of a model is its predictions for observations. Thus the metaphysical
baggage --- the action at a distance, the ether, the caloric, the many
worlds, the objective reality --- should not be taken too
seriously. However, they frequently play a useful pedagogical role.

Models and observations have a symbiotic relation. The atomic nucleus
both shapes the nuclear mean field and is shaped by it. Similarly
observations shape the models and in turn are shaped by them. The most
exalted model can be dethroned by mundane observations while even the
most extraordinary observation is meaningless without the context
provided by the models.

It is often stated by anti-evolution forces that evolution is not a
fact; a rhetorically powerful but ultimately meaningless statement. As
should be obvious from the discussions in this paper, evolution is a
model. A model, by its very nature, never becomes a ``fact'' that is
it never becomes certain but always remains tentative. Trying to
classify evolution or any empirical model as fact or not-fact is a
failure of categories and indicates a profound ignorance of the nature
of empirical knowledge. Evolution is a model, hence tentative, but a
model with extraordinary predictive power. That is high praise, the
highest science can give.  Similar arguments are also made against
other models: science has not proven $X$.  For example $X$ might be
global warming due to green-house gases. Of course science has not
proven $X$. Proofs are the domain of mathematics, not the empirical
sciences. When people use the $X$ is not a fact or $Y$ is not proven
gambits it is a tacit admission they have lost the science argument
and they are just trying to downplay the significance of that failing.

ACKNOWLEDGMENTS: E.D.~Cooper, S.~Coutu, B.S.~Davids, H.W.~Fearing,
 D.~Frekers, S.W.~Hong, M.M.~Pavan, L.~Theu\ss l, E.W.~Vogt and
 R.~Woodside are thanked for reading the manuscript and for useful
 comments. The Natural Sciences and Engineering Research Council of
 Canada is thanked for financial support. TRIUMF receives federal
 funding via a contribution agreement through the National Research
 Council of Canada.


\begin{thebibliography}{99}
\addcontentsline{toc}{part}{References}
\bibitem{Bacon}Novum Organum, Francis Bacon, 1620.
\bibitem{Gopnik} A pilgrimage to Popper, Adam Gopnik, The New Yorker
2002-04-01,\\
\href{http://www.newyorker.com/critics/atlarge/?020401crat\_atlarge}
{http://www.newyorker.com/critics/atlarge/?020401crat\_atlarge}
\bibitem{Hume}An Enquiry Concerning Human Understanding, David Hume,
1748.
\bibitem{Kant}Critique of Pure Reason, Immanuel Kant, 1781
\bibitem{Popper}Conjectures and Refutations; The Growth of Scientific
  Knowledge, Karl Popper, Routledge and Kegan Paul, (London) 1963
\bibitem{Kuhn} The Structure of Scientific Revolutions, Thomas S. Kuhn,
  University Of Chicago Press; 3rd edition (December 15, 1996) 
\bibitem{Wegener}Origin of Continents and Oceans,  Alfred Lothar Wegener, 
Dover Publications (1966); This work was first published in 1915 in German. 
\bibitem{AHD}The American Heritage Dictionary of English Language,
  Third Edition, Houghton Mifflin Company.(Boston)  1992
\bibitem{DoverA}In the United States District Court
for the Middle District Of Pennsylvania, TAMMY KITZMILLER, et
al. v. Dover Area School District; et al, Answer, Point 13.
\href{http://www.pamd.uscourts.gov/kitzmiller/04cv2688-22.pdf}
{http://www.pamd.uscourts.gov/kitzmiller/04cv2688-22.pdf}
\bibitem{Okubo}S.~Okubo, Phys.Lett. B5 (1963) 165; G.~Zweig, CERN
Report No.8419/TH412 (1964); I.~Iizuka, Prog. Theor. Phys. Suppl. 3738
(1966) 21.
\bibitem{Wright}The New Testament and the People of God, N.T.~Wright,
  SPCK (London) 1992 
\bibitem{Polanyi}Personal Knowledge : Towards a Post-Critical
Philosophy. Michael Polanyi, University Of Chicago Press (August 15,
1974)
\bibitem{quantum}See \href{http://musr.physics.ubc.ca/~jess/p200/quotes.html}
{http://musr.physics.ubc.ca/~jess/p200/quotes.html}
for an interesting collection of quotations on quantum mechanics.
\bibitem{Schr}\href{http://www.quotationspage.com/quote/14360.html}
{http://www.quotationspage.com/quote/14360.html}
\bibitem{Einstein}A.~Einstein, B.~Podolsky, and N.~Rosen,
 Phys. Rev. 47, 777-780, 1935.
\bibitem{quantuma}As quoted in ``The philosophy of Niels Bohr'' by
Aage Petersen, Bulletin of the Atomic Scientists (September 1963)
\bibitem{Michelson}A.A.~Michelson and E.W.~Morley, The American
  Journal of Science, 134 (1887) 333.  
\bibitem{Miller}Dayton C.~Miller, Rev. Mod. Phys. 5, (1933) 203.
\bibitem{Brillet}A. Brillet and J. L. Hall, Phys. Rev. Lett. 42, 549
  (1979); D.~Hils and J.L.~Hall, Phys. Rev. Lett. 64, 1697 (1990).
\bibitem{jpsi}J.J.~Aubert, et al, Phys.~Rev.~Lett. 33 (1974) 1404;
  J.-E. Augustin et al. Phys.~Rev.~Lett. 33 (1974) 1406.
\bibitem{Feyerabend}Against Method: Outline of an Anarchistic Theory
  of Knowledge, Paul K Feyerabend, Humanities Press (1975)
\bibitem{Ashmore}M.~Ashmore, Social Studies of Science, 23 (1993), 67.
\bibitem{Franks}Polywater, Felix Franks, The MIT Press, Boston (March
  29, 1983)     
\bibitem{Wietfeldt}F.E.~Wietfeldt and E.B.~Norman, Phys. Rep. 273
(1996) 150
\bibitem{Close}Too Hot to Handle: The Race for Cold Fusion, Frank
  Close, Princeton Univ Pr (April 1991).
\bibitem{Dalton}R.~Dalton, Nature 420  (2002) 6917.
\bibitem{Hicks}M. Battaglieri et el, Phys.Rev.Lett.96:042001,2006.
\bibitem{Laplace}This quote, originally due to Simon Laplace, is
frequently shortened to ``Extraordinary claims require extraordinary
proof''.
\bibitem{pseudo}I.~Langmuir, and R.N.~Hall, Physics Today 42 (10): 36-48. 1989;
 Voodoo Science: The Road from Foolishness to Fraud,  Robert L. Park,
 Oxford U. P., New York, 2000.  
\bibitem{Park}This sentence is quoted form ``The Seven Warning Signs
  of Bogus Science'', Robert L. Park, The Chronicle of Higher
  Education, January 31, 2003,\\
  \href{http://chronicle.com/free/v49/i21/21b02001.htm}
{http://chronicle.com/free/v49/i21/21b02001.htm} 
\bibitem{Shan}R.S.~Shankland, S.W.~McCuskey, F.C.~Leone, and G.~Kuerti
Rev. Mod. Phys. 27, 167 (1955).
\bibitem{Raymond}The Cathedral \& the Bazaar, Eric S. Raymond, O'Reilly. (1999)
\bibitem{Asimov}I. Asimov, The Skeptical Inquirer, 14 (1989) 35;
The Relativity of Wrong, Isaac Asimov, Zebra Books; Reissue edition
(January 1996). 
\bibitem{Poincare}Do an internet search on ``Poincare convention'' for
  many articles on the topic.
\bibitem{Gage}Introduction to Physical Science, Alfred Payson Gage,
  Ginn and Company,  Cambridge, MA (1902) 
\bibitem{Thursday}Last Thursdayism was invented on the USENET group
talk.origins in 1991, as a hyperbolic response to omphalism.
\bibitem{Gosse}Omphalos: An Attempt to Untie the Geological Knot ,
Phillip Gosse, Originally published: London J. Van Voorst 1857, republished
Ox Bow Press; Reprint edition (January 1998).
\bibitem{Everett}H.~Everett, Review of Modern Physics 29, (1957)
454-462;
\bibitem{poincare}The Value of Science (1914), H.~Poincar\'e, Dover
  (New York) 1958, page 14.
\bibitem{science} Teaching About Evolution and the Nature of Science,
Working Group on Teaching Evolution, National Academy of Sciences,
National Academy Press, Washington DC (1998).
\bibitem{NASB}Scripture taken from the NEW AMERICAN STANDARD BIBLE
Copyright 1960,1962,1963,1968,1971,1972,1973,1975,1977,1995 by The
Lockman Foundation. Used by permission.
\bibitem{Vries}P.~de Vries, Christian Scholars Review, 15 (1986) 388.
\bibitem{Benson}H.~Benson et al,  American Heart Journal 151 (2006) 934; 
M.W.~Krucoff, S.W.~Crater and K.L.~Lee,   American Heart Journal 151
(2006) 762.
\bibitem{Drummond}Henry Drummond, The Lowell Lectures on the Ascent of
Man, Glasgow: Hodder and Stoughton, 1904.
\bibitem{Behe} Darwin's Black Box, Michael J. Behe, Touchstone, New
  York, 1996.
\bibitem{Dover}In the United States District Court for the Middle
District Of Pennsylvania, TAMMY KITZMILLER, et al. v. Dover Area
School District; et al, Memorandum of Opinion, Section 4.
\href{http://www.pamd.uscourts.gov/kitzmiller/kitzmiller\_342.pdf}
{http://www.pamd.uscourts.gov/kitzmiller/kitzmiller\_342.pdf}
\bibitem{Wedge}The Wedge Strategy, Center for the Renewal of Science
\& Culture,\\ \href{http://www.antievolution.org/features/wedge.html}
{http://www.antievolution.org/features/wedge.html}
\bibitem{IEEE}There is even a journal dedicated to this field called
``IEEE Transactions on Evolutionary Computation''
\bibitem{Dae}E.B.~Daeschler, N.H.~Shubin and F.A.~Jenkins, Nature, 440 
(2006) 757; ibid 440 (2006) 764.
\end{thebibliography}
\end{document}